\documentclass[pra,aps,preprint,amsmath,amssymb]{revtex4}
\usepackage{graphicx}
\pdfoutput=1
\usepackage{amsmath}
\usepackage{color}
\usepackage{float}
\setcitestyle{super}
\usepackage[english]{babel}
\usepackage{color,soul}
\usepackage{color}

\begin{document}
\title{Tip-induced Superconductivity Coexisting with Preserved Topological Properties in Line-nodal Semimetal ZrSiS}
\author{Leena Aggarwal$^1$}
\author{Chandan K. Singh$^2$}
\author{Mohammad Aslam$^1$}
\author{Ratnadwip Singha$^3$}
\author{Arnab Pariari$^3$}
\author{Sirshendu Gayen$^1$}
\author{Mukul Kabir$^2$}
\author{Prabhat Mandal$^3$}
\email{prabhat.mandal@saha.ac.in}
\author{Goutam Sheet$^1$}
\email {goutam@iisermohali.ac.in}
\affiliation{$^1$Department of Physical Sciences,
Indian Institute of Science Education and Research Mohali,
Sector 81, S. A. S. Nagar, Manauli, PO: 140306, India}

\affiliation{$^2$ Department of Physics and Centre for Energy Science, Indian Institute of Science Education and Research, Pune 411008, India}

\affiliation{$^3$Experimental Condensed Matter Physics Division, Saha Institute of Nuclear Physics, HBNI, 1/AF Bidhannagar, Kolkata 700064, India}

\date{\today}

\begin{abstract}
\textbf{ZrSiS was recently shown to be a new material with topologically non-trivial band structure which exhibits multiple Dirac nodes and a robust linear band dispersion up to an unusually high energy of 2\,eV. Such a robust linear dispersion makes the topological properties of ZrSiS insensitive to perturbations like carrier doping or lattice distortion. Here we show that a novel superconducting phase with a remarkably high $T_c$ of 7.5\,K can be induced in single crystals of ZrSiS by a non-superconducting metallic tip of Ag. From first-principles calculations we show that the observed superconducting phase might originate from  dramatic enhancement of density of states due to the presence of a metallic tip on ZrSiS. Our calculations also show that the emerging tip-induced superconducting phase co-exists with the well preserved topological properties of ZrSiS.} 

\end{abstract}

\maketitle


It is widely believed that inducing superconductivity in topologically non-trivial materials might eventually lead to the discovery of topological superconductors where the elusive particles called Majorana fermions can, in principle, be realized in the vortex state.\cite{Hsu,Lu,Bee,Lei} Motivated by this idea, efforts have been made to induce superconductivity through a number of methods including chemical doping\cite{Wang,Han,Du,Xing} and applying pressure\cite{Zhu,Zhou,Dro, Kong} on topological materials. As a result, superconductivity was observed in a number of topological systems upon chemical doping (e.g., $Cu$-intercalated $Bi_2Se_3$)\cite{Wang} and applying pressure (e.g.,$Bi_2Te_3$)\cite{Kong}. However, such methods worked only on a very limited number of systems. Recently, it was shown that a novel mesoscopic superconducting phase can also be induced in a non-superconducting topological system simply by making a point contact with sharp needle made of a simple elemental normal metal.\cite{Sheet1,Sheet2} The observed supercondcuting phase originating only under a point of contact is known as tip-induced superconductivity (TISC). This discovery expanded the landscape over which the possibility of a topological superconductor can be explored. 

In this paper, we report the discovery of TISC in a novel topological material ZrSiS.\cite{Xu, Neu} The TISC on ZrSiS shows a critical temperature ($T_c$) of $\sim$ 7.5\,K and a superconducting energy gap $\Delta_{(T = 0)} \sim$ 1\,meV. Such discovery is particularly important because ZrSiS is known to show a linear band dispersion over a large energy range $\sim$ 2\,eV\cite{Les} whereas the linear extent for most of the other topological materials is limited up to few hundred meV from respective Dirac points.\cite{Jin, Ko} This makes the topological properties of ZrSiS extremely robust against carrier doping, variation of stochiometry and other external perturbative effects thereby implying that when a superconducting phase is realized on ZrSiS point contacts, the topological properties of ZrSiS are not expected to be destroyed merely by the presence of a metallic tip forming the point contacts. Therefore, the superconducting phase realized on ZrSiS does not emerge at the expense of the topological nature of ZrSiS. This makes ZrSiS the most promising candidate for a topological superconductor and the idea is also supported by our first-principles calculations.


The ZrSiS single crystals were grown in standard iodine vapor transport method. The polycrystalline powder was prepared in two steps. At first, elemental Si (Strem Chem. 99.999\%) and S (Alfa Aesar 99.9995\%) were mixed in stoichiometric ratio (1:1) and heated at 1000$^{o}$C under vacuum. The resultant powder was mixed with elemental Zr (Alfa Aesar 99.9\%) and again heated at 1100$^{o}$C. This polycrystalline ZrSiS along with iodine (5mg/cm$^3$) were sealed in a quartz tube under vacuum and kept in a gradient furnace for 72\,h. During this period, a temperature gradient of 100$^{o}$C was maintained along the quartz tube with hotter end at 1100$^{o}$C. Shinny rectangular plate like crystals were obtained at the cooler end. The single crystals have been characterized by high-resolution transmission electron microscopy (HRTEM) and energy-dispersive X-ray (EDX) spectroscopy in a FEI, TECNAI G2 F30, S-TWIN microscope operating at 300\,kV. The details of characterization techniques are described elsewhere.\cite{Mandal}

\begin{figure}[h!]
\centering
	\includegraphics[width=0.8\textwidth]{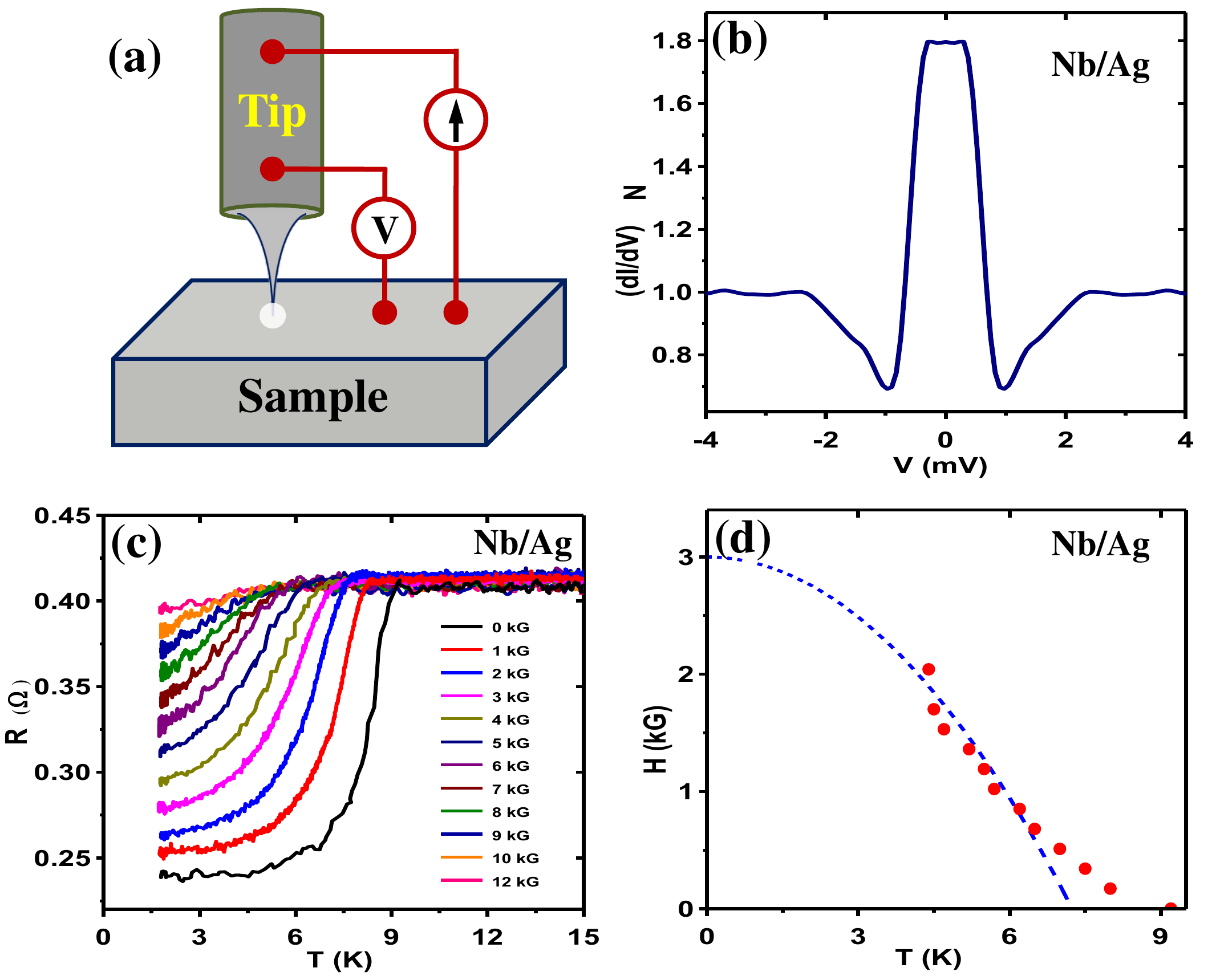}
	\caption{Point-contact spectroscopy on a conventional superconductor, Nb using Ag tip. (a) A schematic of a point-contact spectroscopy experiment. (b) A normalized diffrential conductance (($dI/dV)_{N}$) \emph{vs} applied bias ($V$) point contact spectrum in thermal regime of transport for a conventional superconductor, Nb using Ag tip at 1.8\,K. (c) Resistance versus temperautue ($R-T$) curves obtained for Nb/Ag point-contact showing systematic drop of the superconducting transition temperature with varying the applied magnetic field. (d) The $H-T$ phase diagram extracted from Figure 1(c) in blue dashed line is empirically expected and red dots are experimental data.}
	\label{f1}
\end{figure}


Figure 1 (a) depicts a schematic of the point contacts. The point contacts were made by bringing sharp metallic tips of silver (Ag) on the surface of single crystals of ZrSiS (Also see Figure S1 in the supplementary materials). The experiments were performed on two independently grown single crystals in order to confirm the reproducibility of the results (see Figure S3, S4, S5 in supplementary materials). A lock-in based modulation technique was used to measure the differential conductance, $dI/dV$ of the point-conatcts as a function of an applied dc bias, $V$ (see Figure S2 in supplementary materials).

\begin{figure}[h!]
\centering
		\includegraphics[width=0.7\textwidth]{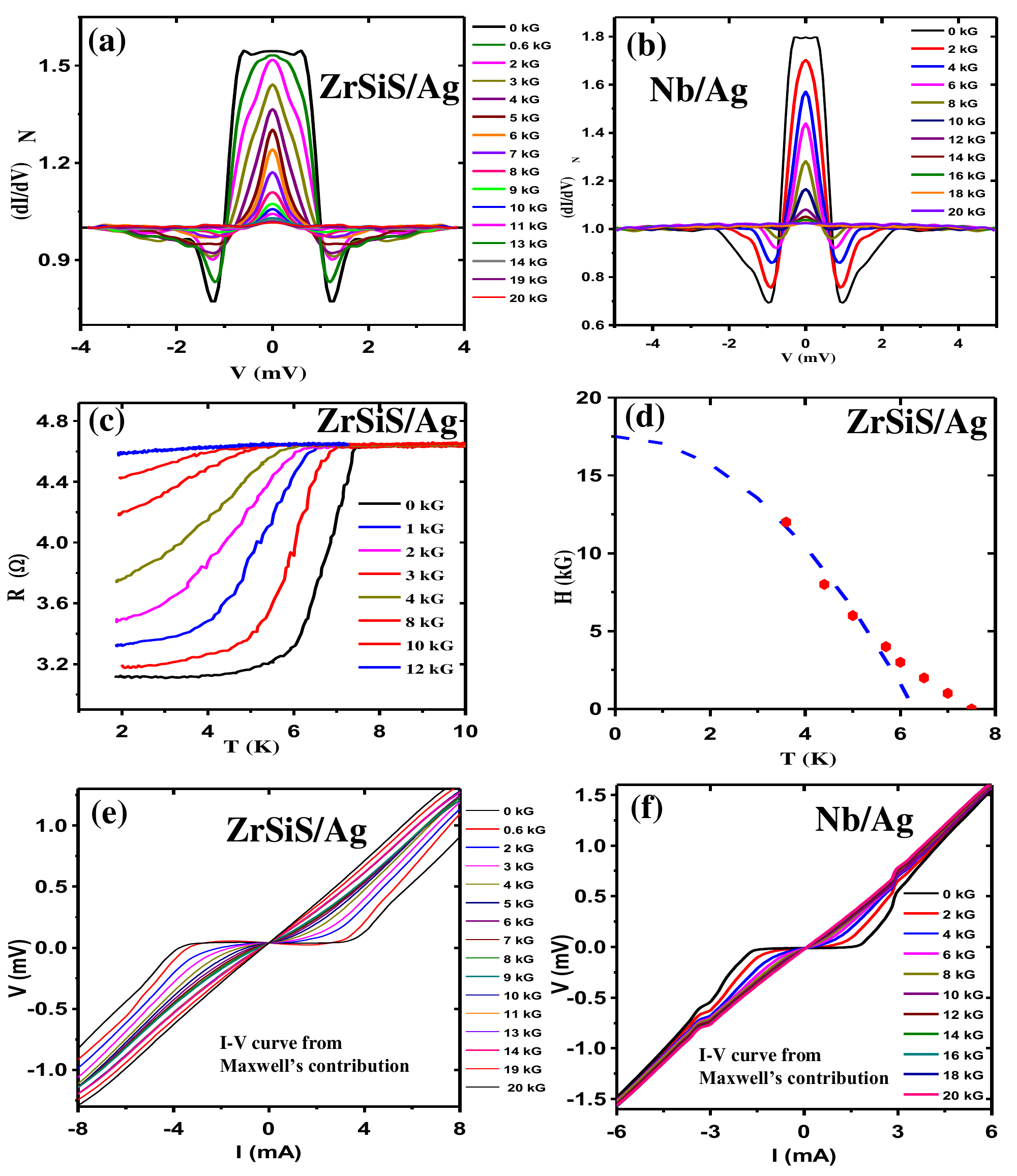}
	\caption{A comparision of $(dI/dV)_{N}$ \emph{vs} $V$ spectra on ZrSiS/Ag point-contact with Nb/Ag point-contact spectra in thermal regime of transport by showing magnetic field dependent data. Magnetic field dependent point-contact spectra at 1.8\,K in thermal regime of transport (a) on ZrSiS/Ag point-contact, (b) on Nb/Ag point-contact. (c) Magnetic field dependent $R-T$ curves showing disapearance of superconducting transition temperature with increasing magnetic field for ZrSiS/Ag point-contact. (d) The $H-T$ phase diagram extracted from Figure 2(c) in blue dashed line is empirically expected and red dots are experimental data. Magnetic field dependence of $I-V$ curves at 1.8\,K calculated by BTK theory containing Maxwell's contribution only (e) for ZrSiS/Ag point-contact, (f) for Nb/Ag point-contact.}
	\label{f2}
\end{figure}

Usually, for mesoscopic point contacts, the resistance $R_{PC}$ is given by Wexler's formula\cite{Wexler}: $R_{PC} = \frac{2h/e^2}{(ak_F)^2} + \Gamma (l/a)\frac{\rho (T)}{2a}$, where $h$ is Planck's constant, $e$ is the charge of a single electron, $a$ is the contact diameter, $k_{F}$ is a Fermi momentum, $\Gamma(l/a)$ is a slowly varying function of the order of unity, $\rho$ is the bulk resistivity of the material and $T$ is the effective temperature at the point-contact. The first term is known as ballistic or Sharvin's resistance ($R_{S}$)\cite{Supriya} that depends on the fundamental constants $h$, $e$ and the number of conducting channels in the point contacts. The second term in Wexler's formula is called the Maxwell's resistance ($R_{M}$) which is qualitatively similar to the bulk resistance and depends directly on the resistivity of the materials forming the point contact. The Wexler's formula also suggests that when the contact diameter is small compared to electronic mean free path \emph{i.e.}, when the contact is in the so-called ballistic regime, $R_{S}$ dominates and in the other extreme called the thermal regime, $R_{M}$ contributes most to $R_{PC}$. Therefore resistive transitions lead to non-linearities in the $I-V$ characteristics corresponding to $R_{M}$ of point-contacts.\cite{Goutam} For superconducting point contacts in the thermal regime, as the current through a point contact goes down below the critical current ($I_c$) associated with the point contact, the resistivity ($\rho$) of the superconducting component of the point contact will be zero leading to a sharp change in $R_{M}$.\cite{Goutam,Naidyuk} This change introduces a large non-linearity in the $I-V$ characteristic of a thermal regime point contact, resulting in two sharp dips in the differential conductance ($dI/dV$) \emph{vs} dc bias ($V$) spectra. Such dips are seen to be symmetric about $V = 0$ and the position of the dips normalized by the normal state resistance of the point contacts provides a direct estimate of the magnitude of $I_c$.

In order to highlight the points discussed above directly, in Figure 1(b), we show a $(dI/dV)_N$ \emph{vs} $V$ spectrum acquired in the thermal regime point contact between a known conventional superconductor niobium (Nb) and a sharp metallic tip of silver (Ag). The conductance dips symmetric about $V = 0$ are clearly visible. Furthermore, as shown in Figure 1(c), the resistance ($R$) vs. temperature ($T$) of a superconducting (thermal regime) point contact on Nb shows a sharp change corresponding to the superconducting transition of Nb at 9.2\,K. The $R-T$ evolves systematically involving a monotonic decrease in $T_c$ with increasing magnetic field ($H$).

In the $R-T$ data obtained from Nb point contacts, it is clear that the resistance does not become zero down to 1.4\,K. This shows that for a superconducting point-contact, the contact resistance does not become zero. This is primarily due to several reasons including (i) the existence of a non-superconducting component of the point-contact, (ii) a small contribution from Sharvin's resistance that might be present  even when the contacts are far from the ballistic regime and (iii) possible mismatch of the Fermi-velocities in the two materials forming the point-contact.\cite{Goutam_thesis} Hence the measurement of ``zero-resistance" is not a hallmark signature of superconductivity in a superconducting point contact. However, despite the absence of a zero-resistance in the measurement, from other hallmarks of supercondcutivity like, the field dependent critical current, combined with the field dependent $R-T$ data, a mesoscopic tip-induced superconducting (TISC) phase emerging only under a point contact can be detected and verified. Figure 1(d) shows the $H-T$ phase diagram (red dots) extracted from magnetic field dependence $R-T$ data as shown in Figure 1(c). The blue dashed line in the same Figure is expected $H-T$ phase diagram for a conventional superconductor as known empirically.

\begin{figure}[h!]
\centering
		\includegraphics[width=0.8\textwidth]{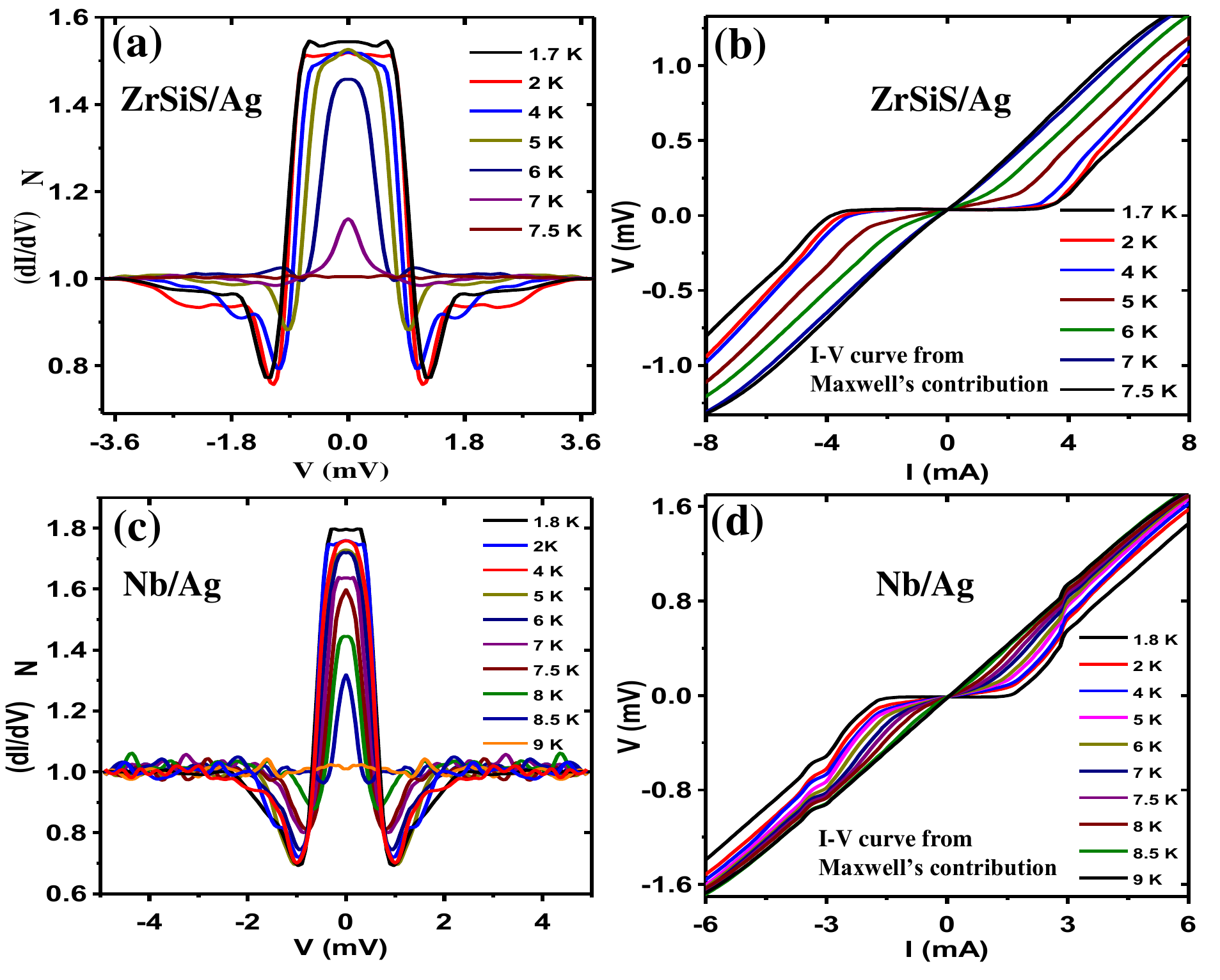}
	\caption{A comparision of $(dI/dV)_{N}$ \emph{vs} $V$ spectra on ZrSiS/Ag point-contact with Nb/Ag point-contact spectra in thermal regime of transport by showing temperature dependent data. (a) Temperature dependence of thermal limit spectra on ZrSis/Ag point-contact. (b) Temperature dependence of $I-V$ curves calculated by BTK theory containing Maxwell's contribution only for ZrSiS/Ag point-contact. (c) Temperature dependence of the thermal limit spectra on Nb/Ag point-contact (d) Temperature dependence of $I-V$ curves calculated by BTK theory containing Maxwell's contribution only for Nb/Ag point-contact.}
	\label{f2}
\end{figure}

Now we focus on the data obtained on high quality single crystals of ZrSiS. In Figure 2(a) we present the magnetic field dependence of a spectrum obtained on ZrSiS/Ag point contacts. The striking similarity of the zero field data with that obtained on Nb point contacts must be noted. The conductance dips appearing in the spectrum originate from the critical current of the point contact. All the spectral features show monotonic evolution with increasing magnetic field until all the features smoothly disappear at 20\,kG. In order to highlight the superconducting nature of ZrSiS/Ag point contacts, in Figure 2(b) we provide the systematic magnetic field dependent data obtained on supercondcuting Nb/Ag point contacts. The qualitative similarity in the magnetic field dependence of ZrSiS point contacts and Nb point contacts is clear. 

In Figure 2(c) we present the magneto-transport data on ZrSiS/Ag point contacts. The superconducting transition temperature ($T_C$) is seen to be 7.5\,K. This value is remarkably high and comparable to the $T_C$ of the celebrated elemental superconductors like Pb and Nb. The critical temperature decreases monotonically with increasing magnetic field. Again, the zero resistance state cannot be directly measured for reasons discussed before. The $H-T$ phase diagram extracted from these data has been shown in Figure 2(d). The blue dashed line in Figure 2(d) shows the expected dependence for a conventional superconductor as known empirically. The projected upper critical field of the supercondcuting phase could be as high as 17\,kG and our experimental data points (red dots) deviate slighty from the empirical expectation. This might be an indication of the existence of an unconventional component in the supercodncuting phase realized on the topologically non-trivial system ZrSiS.

In order to illustrate the role of critical current on the point contact spectra obtained on ZrSiS, we have used BTK theory\cite{BTK} to calculate an approximate $I-V$ characteristic associated with the ballistic component ($R_{S}$) of the point contact resistance and then subtracted the same from the $I-V$ associated with the total point contact resistance ($R_{PC}$). The resultant $I-V$ thus obtained is associated with the Maxwell's contribution ($R_{M}$) alone. The resultant $I-V$ characteristic and the magnetic field dependence of the same has been shown in Figure 2(e). This $I-V$ directly reflects the non-linearities associated with the critical current of the superconducting point contact. The smooth decrease of the range over which the voltage remains zero with increasing magnetic field is fully consistent with the magnetic field dependence of a critical current dominated $I-V$ for a superconducting point contact. For comparison, in Figure 2(f) we have also provided similar data extracted from the magnetic field dependent point contact spectra obtained on superconducting Nb (Figure 2(b)).

To provide further support for the superconducting origin of the spectra presented here, we have performed detailed temperature dependence of the spectra. In Figure 3(a) we show the temperature dependence. The spectral features disappear at the critical temperature. We have also extracted the temperature evolution of the critical current dominated $I-V$ following the same protocol discussed above (Figure 3(b)). For comparison, we have also included the temperature dependent data on superconducting Nb point contacts (Figure 3(c,\,d)). 

\begin{figure}[h!]
		\includegraphics[width=0.8\textwidth]{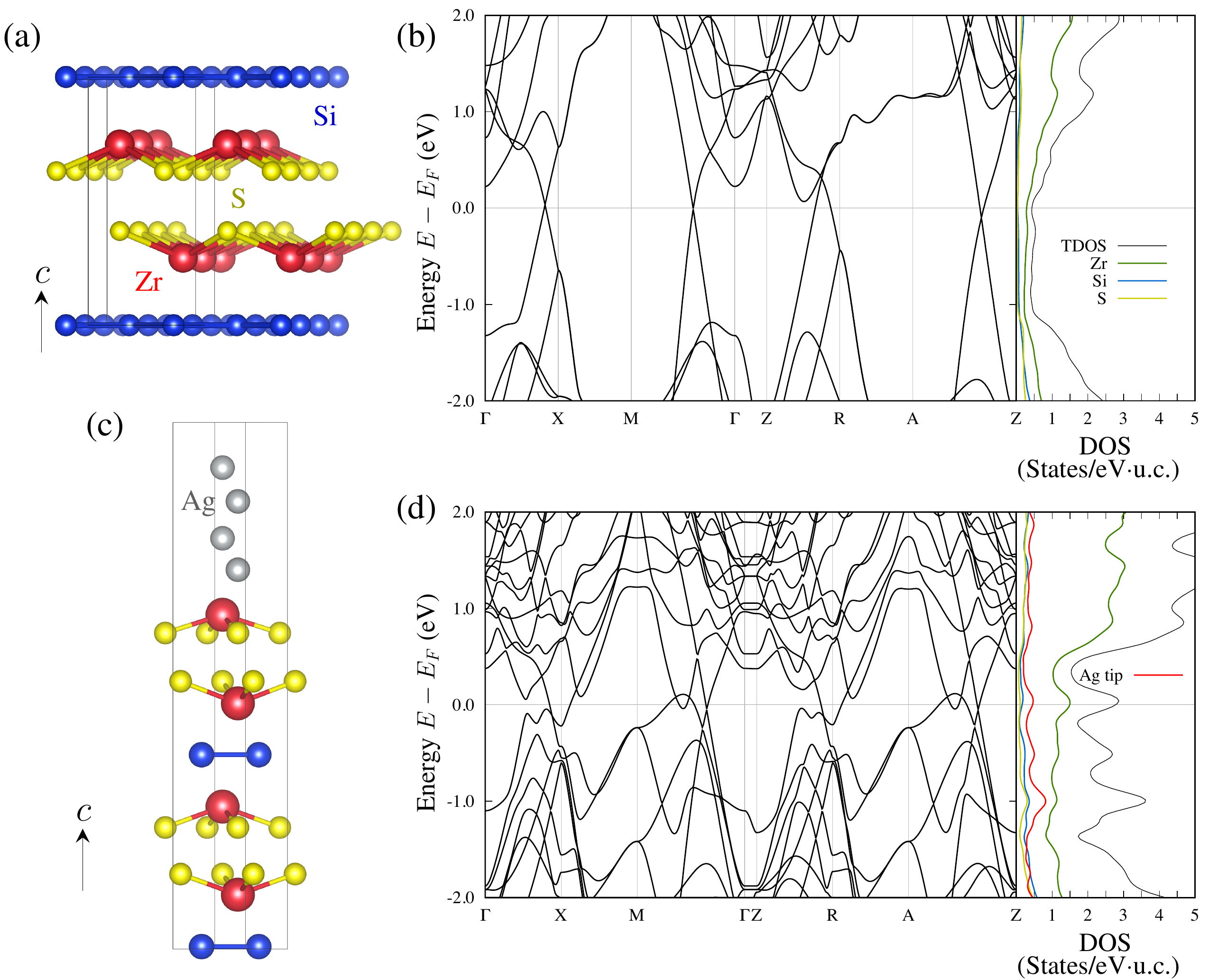}
\caption{The crystal lattice and band structure of ZrSiS without and with Ag-tip. (a) ZrSiS crystallizes in tetragonal $P4/nmm$ structure. The Zr and S layers are sandwiched between the Si square nets in a fashion that the two neighbouring S layers are between the Zr layers. Calculated lattice parameters are in excellent agreement with experimental data.\cite{Les} (b) Calculated band structure of ZrSiS without spin-orbit coupling showing Dirac cones along $\Gamma$X, M$\Gamma$ and AZ lines of the Brillouin zone. The corresponding DOS indicate Zr-$d$ character around the Fermi level. (c) The 1$\times$1$\times$3 [ZrSiS]$_4$Ag$_4$ superlattice to investigate the effect of Ag-tip. (d) The superlattice band structure, showing Dirac cones along the M$\Gamma$ and AZ lines. Resulting DOS at the $E_F$ show substantial increase.}

	\label{f2}
	
\end{figure}

Even when a point contact is in the thermal regime of transport, as per Wexler's formula\cite{Wexler}, a small but finite ballistic component exists. From a visual inspection it is clear that between -0.8 meV and +0.8 meV, the zero field spectrum on ZrSiS remains flat. This could be a signature of Andreev reflection when the barrier potential at the interface is extremely small making the barriers almost transparent. Such an attribution is consistent with the prediction from the theory of Blonder, Tinkham and Klapwijk (BTK)\cite{BTK} for superconducting point contacts. Going by the arguments provided by BTK, there is a strong indication that the superconducting energy gap ($\Delta$) of the superconducting phase realized on ZrSiS is 0.8 meV. 



In order to understand the mechanism through which superconductivity might be induced in the system, we have carried out first-principles calculations and investigated the change in Fermi surface properties of ZrSiS due to the presence of a metallic tip. Our first-principles calculations are based on density functional theory, within the projector augmented wave formalism.\cite{DFT1, DFT2, PAW} The wave functions were expanded in the plane wave basis with 400\,eV cutoff. The structures were fully optimized using 31$\times$31$\times$19 Monkhorst-Pack $k$-grid to sample the Brillouin zone. The Perdew-Burke-Ernzerhof functional form of generalized gradient approximation was used to describe the exchange-correlation energy.\cite{PBE} Calculated lattice parameters for bulk $P4/nmm$ ZrSiS (Figure 4(a), $a$ = 3.55 and $c$ = 8.14 \AA) are in excellent agreement with the experimental results.\cite{Les} The electronic structure of bulk ZrSiS hosts several Dirac cones along $\Gamma$X, M$\Gamma$ and AZ lines (Figure 4(b)). However, due to the $C_{2v}$ symmetry lines, the Dirac cones at the Fermi level open up small gap $\sim$ 35 meV, while spin-orbit coupling is considered (see Figure S6 in supplementray materials). The density of states (DOS) is found to be 0.86 states/eV$\cdot$unit-cell at $E_{F}$, which is dominated by Zr-$d$ character.  To investigate the effects of metallic Ag point contacts on the electronic structure of ZrSiS, we have considered 1$\times$1$\times$3 [ZrSiS]$_4$Ag$_4$ (shown in Figure 4(c)) and 3$\times$1$\times$1 [ZrSiS]$_4$Ag$_6$ superlattice (presented in Figure S7 in the supplementray report).  The Ag-tip substantially perturbs the band structure of ZrSiS and for both cases, we observe enormous increase in DOS at the $E_{F}$, 2.77 and 3.73 states/eV$\cdot$unit-cell, respectively. In both cases, the contribution of Ag in terms of density of states at and around the Fermi level is considerably small. Thus, it can be surmised that substantial increase in the carrier density around the $E_{F}$ has explicit role in observed tip-induced superconductivity in ZrSiS. It is important to note that while the Dirac cone along the $\Gamma$X disappears due to Ag-tip insertion for [ZrSiS]$_4$Ag$_4$ superstructure,  the Dirac cones along the M$\Gamma$ and AZ lines are found to be protected along with the appearance  of extra nesting feature (Figure 4(d)). This observation hints to the co-existience of topological character of ZrSiS along with tip-induced superconductivity thereby opening a strong possibility of topological superconductivity at ZrSiS/Ag point contact interfaces.

It is true that in point contact spectroscopy, some amount of pressure is applied underneath the tips. Hence, it is also important to look at the possibility of a pressure induced superconducting phase in ZrSiS. However, first of all, a high pressure superconducting phase of ZrSiS is not known to exist. In the high-pressure transport measurements on ZrSiS in our lab we did not find any signature of superconductivity in the system up to 8 GPa. Furthermore, by including the effect of strain in our calculations, did not find any significant change in the fermi surface density of states. Therefore, it is most likely that the exotic superconducting phase emerging under the point contacts on ZrSiS do not arise from pressure. 
     

In the light of our experimental and theoretical investigations it might be rational to surmise that pristine ZrSiS might be on the verge of a superconducting ground state, but some intrinsic processes like order parameter fluctuations due to the fragility of phase coherence destabilize such a ground state. In the past, it was shown that a non-superconducting system possesing a phase-fluctuating supercodnucting order parameter might be stabilized into a supercondcuting ground state simply by increasing the stiffness of the fluctuating phase of the complex superconducting order parameter through an enhancement of superfluid density.\cite{aslam} Such an idea is consistent with our observation of tip-induced superconductivity in ZrSiS, where the sole role of the tip is to increase the local density of states leading to an increase in local superfluid density. 

In conclusion, we have shown that superconductivity can be induced in the topological semimetal ZrSiS using a metallic tip. The transition temperature of such a phase is very high (7.5\,K). Our first principles calculations hint to the possibility of coexistence of topological properties and superconductivity under ZrSiS point contacts thereby making the tip-induced supercondcuting phase a strong candidate for topological supercondcutivity.

\textbf{Acknowledgement:} G.S. would like to acknowledge partial financial support from a research grant from DST-Nanomission under the grant number SR/NM/NS-1249/2013 and from SERB under the grant number EMR/2015/001650. Calculations were done using the supercomputing facilities at the Centre for Development of Advanced Computing, Pune; Inter University Accelerator Centre, Delhi. M.K. acknowledges the funding from the Department of Science and Technology, Government of India, under Ramanujan Fellowship and Nano Mission project SR/NM/TP-13/2016.

\newpage

\begin{center}
	\Large{\textbf{\underline{Supplementary materials}}}
\end{center}

\begin{center}
	\textbf{\underline{Point-contact Spectroscopy}}
\end{center}

Point-contact spectroscopy experiments were performed using a home-built low-temperature probe. The probe consists of a long stainless steel tube at the end of which the probe-head is mounted. The probe head is equipped with a 100 threads per inch (\emph{t.p.i.}) differential screw that is rotated by a shaft running to the top of the cryostat. The screw drives a tip-holder up and down with respect to the sample. The sample-holder is a circular copper disc of diameter 1". A cernox thermometer was mounted on the copper disc for the measurement of the temperature. The temperature of the disc was varied by a heater mounted on the same copper disc (as shown in Figure S1).

\begin{figure}[h!]
	\centering
		\includegraphics[width=0.6\textwidth]{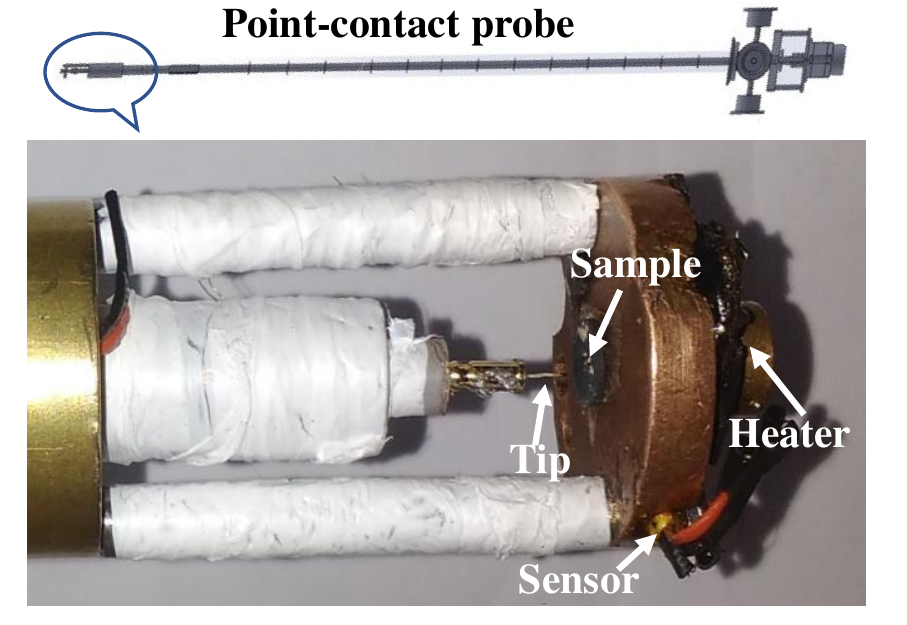}
	\caption{\textbf{ The real image of the point-contact probe:} A close view of the bottom part of point-contact probe showing tip-sample assembly.}
	\label{f2}
	
\end{figure}

The tips were fabricated by cutting the metal wire (diameter 250\,$\mu$m) at an angle. The tip was mounted on the tip holder and two gold contact leads were made on the tip with silver epoxy. The samples were mounted on the sample holder and two silver-epoxy contact leads were mounted on the sample as well. These four leads were used to measure the differential resistance ($dV/dI$) across the point-contacts. 

The point-contact spectra were captured by ac-modulation technique using a lock-in-amplifier (Model: SR830 DSP). A voltage to current converter was fabricated to which a dc input coupled with a very small ac input was fed. The output current had a dc and a small ac component. This current passed through the point-contact. The dc output voltage across the point-contact, $V$ was measured by a digital multimeter (model: Keithley 2000) and the ac output voltage was measured by a lock-in amplifier working at 721\,Hz. The first harmonic response of the lock-in could be taken to be proportional to the differential change in the voltage $dV/dI$.  Properly normalized $dI/dV$ is plotted against $V$ to generate the point-contact spectrum. The software for data acquisition was developed in house using lab-view.

\begin{figure}[h!]
	\centering
		\includegraphics[width=0.6\textwidth]{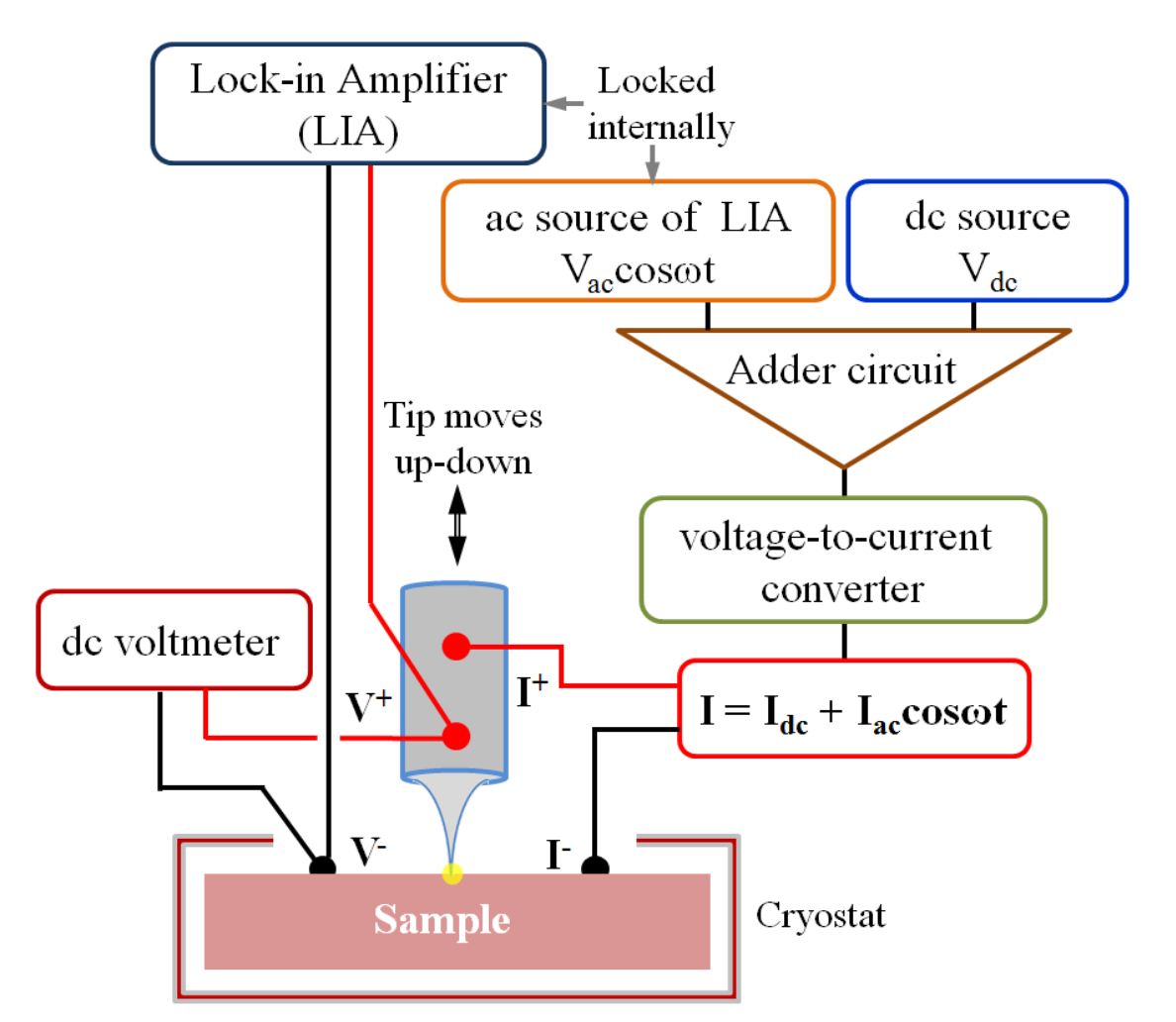}
	\caption{\textbf{Experimental arrangement of point-contact spectroscopy measurements:} The schematic diagram of electrical connections for point-contact measurements. It describes a lock-in based modulation technique which is used to record the diffential resistance $(dV/dI)$ (using a lock-in amplifier, SR 830) with respect to the dc voltage drop across the junction (measured by a digital multimeter, Kiethley 2000). The data is normalized to obtain the PC spectrum  \emph{i.e.} $(dI/dV)_N$ \emph{vs} $V$.}
	\label{f2}
	
\end{figure}

\begin{center}
	\textbf{\underline{How did we determine the critical temperature ($T_c$)?}}
\end{center}

For all the point-contacts reported here we have measured the temperature ($T$) dependence of the point-contact resistance ($R$) with V\,=\,0. The R\,-\,T data show a broad transition to the superconducting state. We have drawn the slope of the R\,-\,T curves above and below the onset of transitions. The temperature at which the two slopes for a given R\,-\,T curve meet has been taken as the $T_c$ for the corresponding point-contacts. It is important to note that for the ballistic point-contacts we cannot measure the $T_c$ as for such point-contacts the contact-resistance depends only on fundamental constants and remain temperature independent. However, from the thermal limit point-contacts we learn that the $T_c$ does not have a strong dependence on contact size for a given sample.


\begin{center}
\textbf{\underline{Determination of the $I-V$ characteristics corresponding to $R_{M}$}} 
\end{center}

For a point-contact between two different materials, the resistance is given by Wexler's formula\cite{Wexler} that has contribution from both ballistic or Sharvin's resistance ($R_{S} = \frac{2h/e^2}{(ak_F)^2}$) and thermal or Maxwell's resistance ($R_M = \frac{\rho (T)}{2a}$). $R_{S}$ is always finite and depends only on fundamental constants namely the Planck's constant ($h$) and the charge of a single electron ($e$). $R_{M}$ is directly dependent on the resistance of the materials forming the point-contact and therefore becomes zero in the superconducting state. The measured data on the point-contacts have contribution of both $R_{M}$ and $R_{S}$. As per Wexler's formula, the total point-contact resistance is $R = R_{S} + R_{M}$. In order to extract the $I-V$ corresponding to the $R_{M}$ component alone we have subtracted the ballistic component ($R_{S}$) from the total resistance ($R$). The ballistic $I-V$ for superconducting point-contacts is dominated by Andreev reflection. The Andreev reflection dominated ballistic $I-V$ was calculated as discussed below using standard BTK theory\cite{BTK} for superconducting point-contacts and the data before and after subtraction are presented in Figure 2 (e,f) and 3 (b,d) of the manuscript.

As per BTK theory, the $I-V$ characteristics of a superconducting point-contact can be generated by using the expression of the current through a ballistic interface given by 

$I_{ballistic}$= $ C \int^{+\infty}_{-\infty} [f(E-eV)-f(E)][1+A(E)-B(E)] dE \, $

where, $A(E)$ is the Andreev reflection probability and $B(E)$ is the normal reflection probability.
     
$A(E)$ and $B(E)$ were calculated using the following formula:

Since the superconducting phase is found to be unconventional, in order to make the analysis more general, for our simulation we have used the modified BTK formula that also accounts for finite quasiparticle lifetime ($\Gamma$) at the interface as described by Plecenik et.al.\cite{plecenik}:
   
 $A(E) = aa^*$ and $B(E) = bb^*$, where the coefficients $a$ and $b$ are given by
       
       $a= u_{0}v_{0}/\gamma,
       b= -(u_{0}^{2}-v_{0}^{2})(Z^{2}+ \iota Z)/\gamma$, where $u_{0}^2$ and $v_{0}^2$ are the probabilities of an electronic state being occupied and unoccupied respectively:

       $u_{0}^{2}= \frac{1}{2}[1 + \frac{\sqrt{(E+ \iota \varGamma)-\Delta^2}}{E+\iota \varGamma}],
       v_{0}^{2}=1-u_{0}^{2}    
       \gamma^{2}=\gamma\gamma^\ast,
       \gamma= u_{0}^{2}+ (u_{0}^{2}-v_{0}^{2})Z^{2}$\\
     
$Z$ is the dimensionless parameter used in BTK theory and it is directly proportional to the strength of the potential barrier at the point-contact interface. For all our simulation the value of $\Gamma$ remained zero. The constant $C$ has been determined by matching the scales of the experimental data and the theoretical curves.  

\begin{center}
	\textbf{\underline{Additional experimental data}}
\end{center}   

\begin{figure}[h!]
	\centering
		\includegraphics[width=0.8\textwidth]{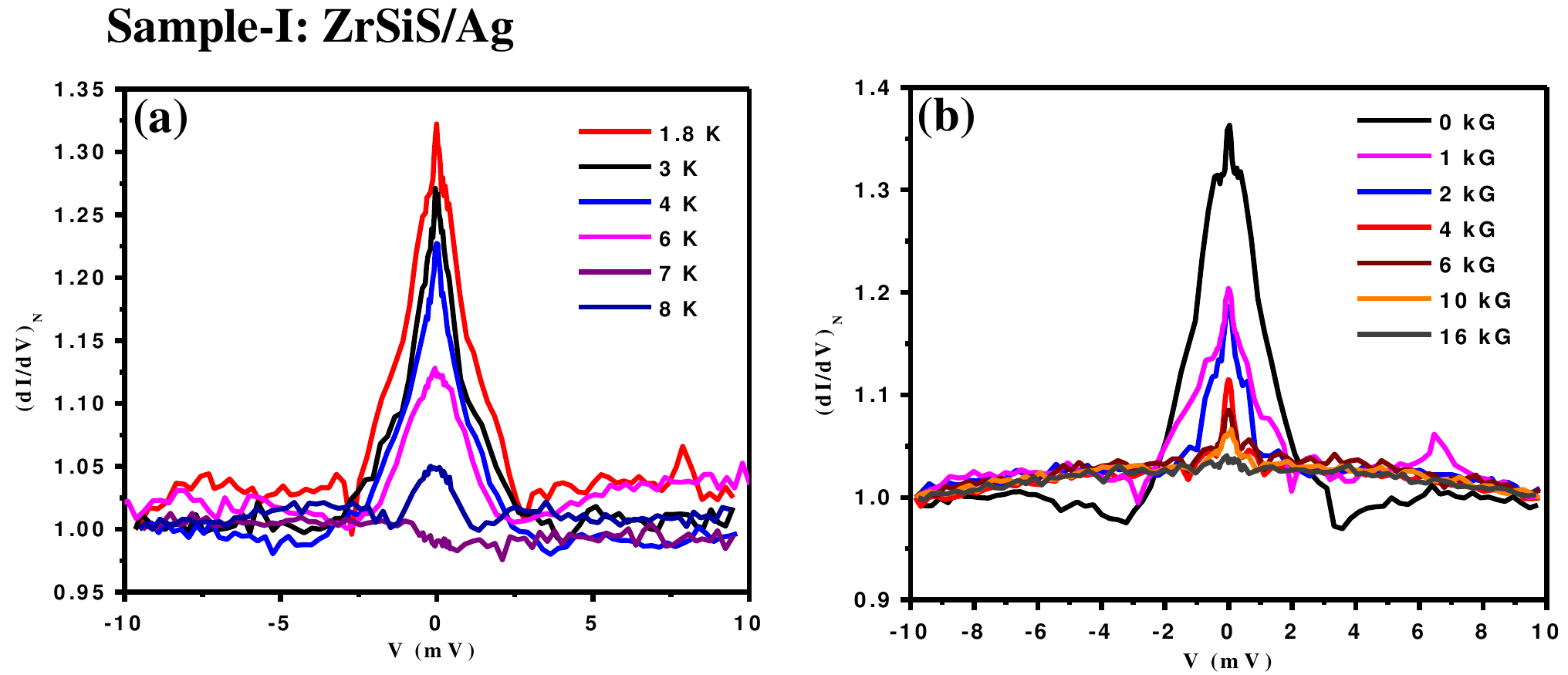}
	\caption{\textbf{$(dI/dV)_{N}$ vs. $V$ spectra on ZrSiS/Ag at different point of contact on the same Sample-I as discussed in the main manuscript :} (a) Magnetic field dependent point-contact spectra at 1.8\,K in thermal regime of transport on ZrSiS/Ag point-contact (b) Temperature dependent point-contact spectra at 1.8\,K in thermal regime of transport on ZrSiS/Ag point-contact. This spectrum was obtained far away from the ballistic regime and consequently the flat region at low bias is absent.}
	\label{f2}
 	
\end{figure}

\newpage
\begin{figure}[h!]
	\centering
		\includegraphics[width=0.6\textwidth]{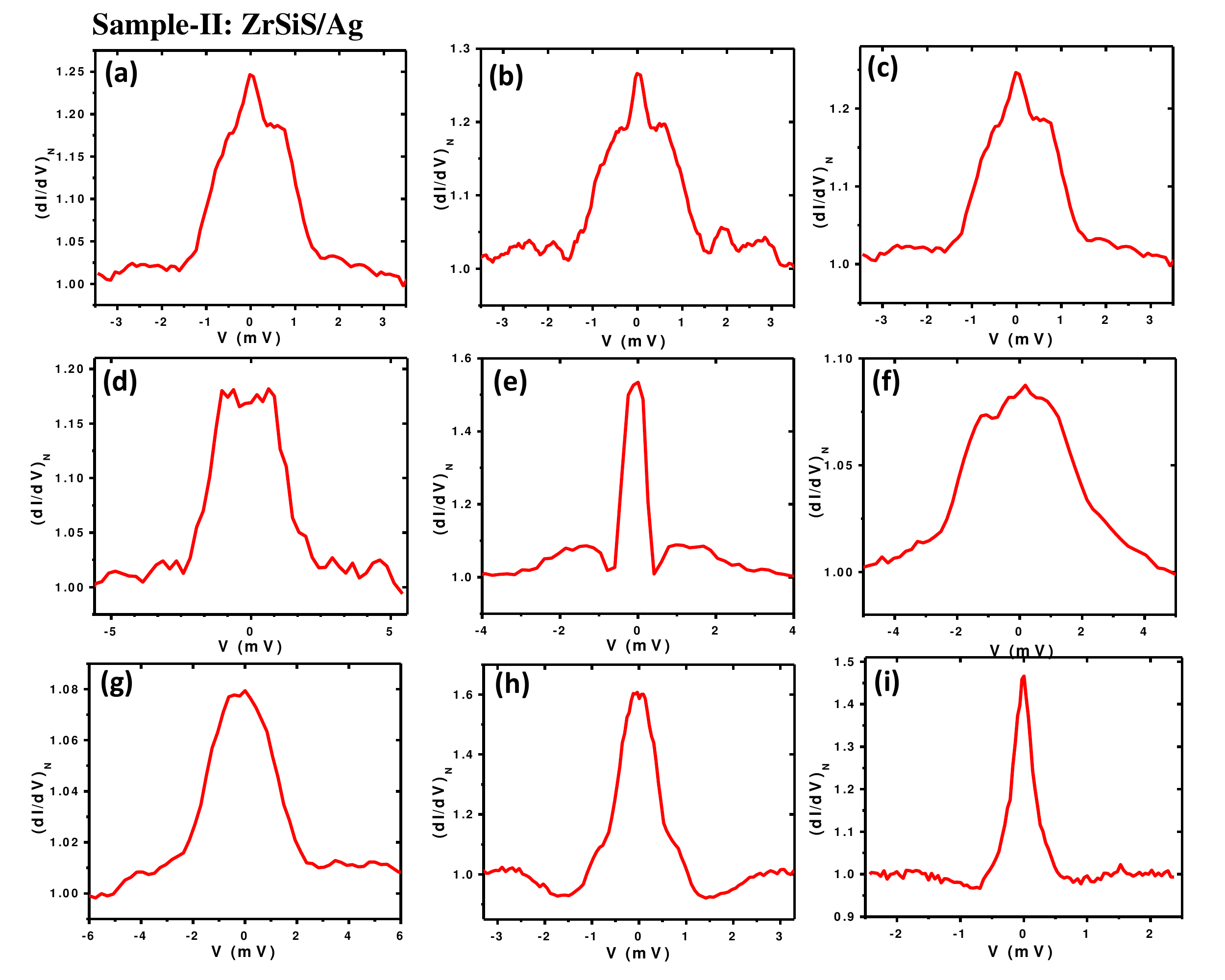}
	\caption{$(dI/dV)_{N}$ vs. $V$ spectra at different points of contact on another sample of ZrSiS named as ``Sample-II" using Ag tip.}
	\label{f2}
	
\end{figure}

\begin{figure}[h!]
	\centering
		\includegraphics[width=0.6\textwidth]{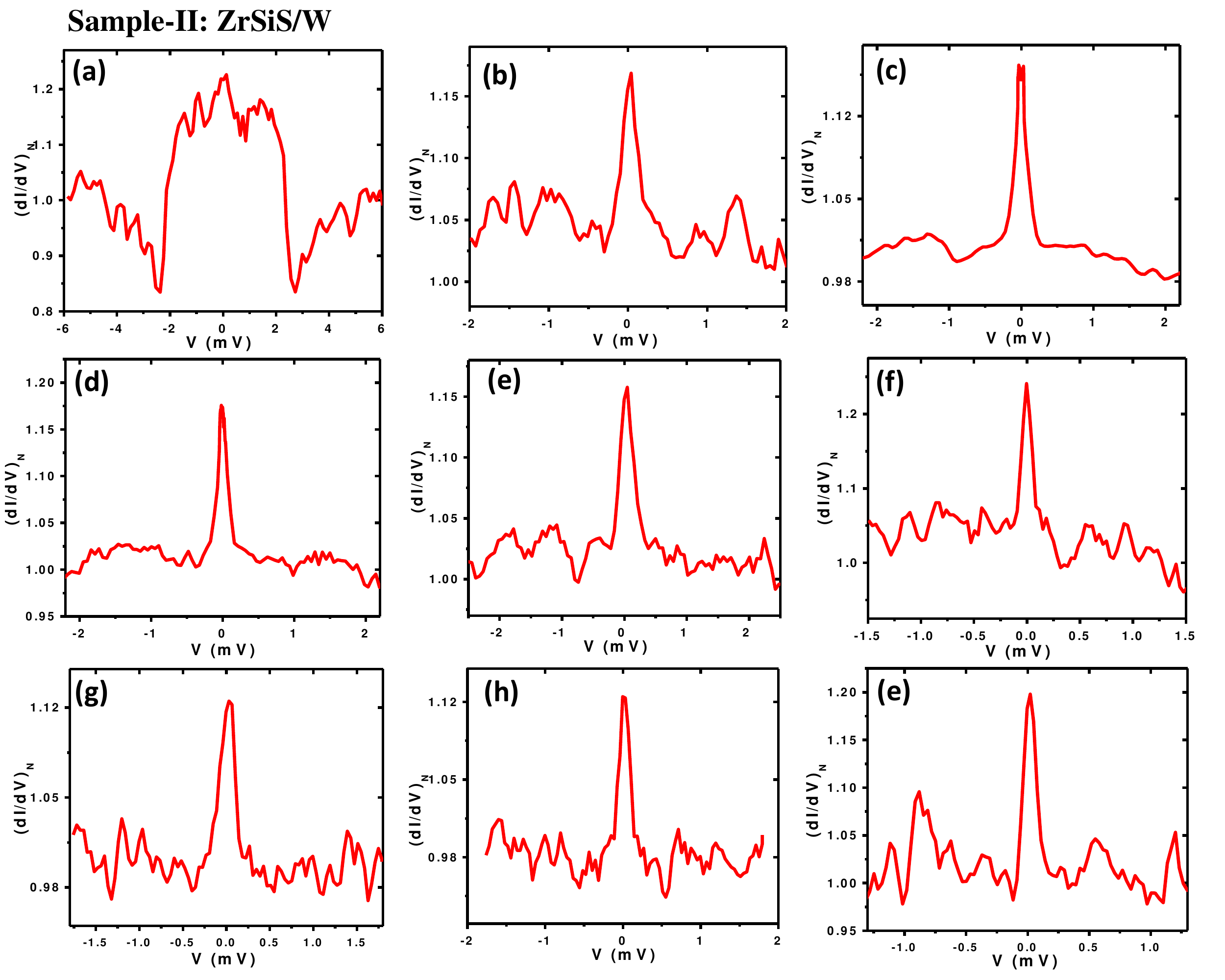}
	\caption{$(dI/dV)_{N}$ vs. $V$ spectra at different points of contact on another sample of ZrSiS named as ``Sample-II" using tungsten (W) tip.}
	\label{f2}
	
\end{figure}

\newpage

\begin{center}
	\textbf{\underline{Additional theoretical data}}
\end{center}

\begin{figure}[h!]
	\centering
		\includegraphics[width=0.7\textwidth]{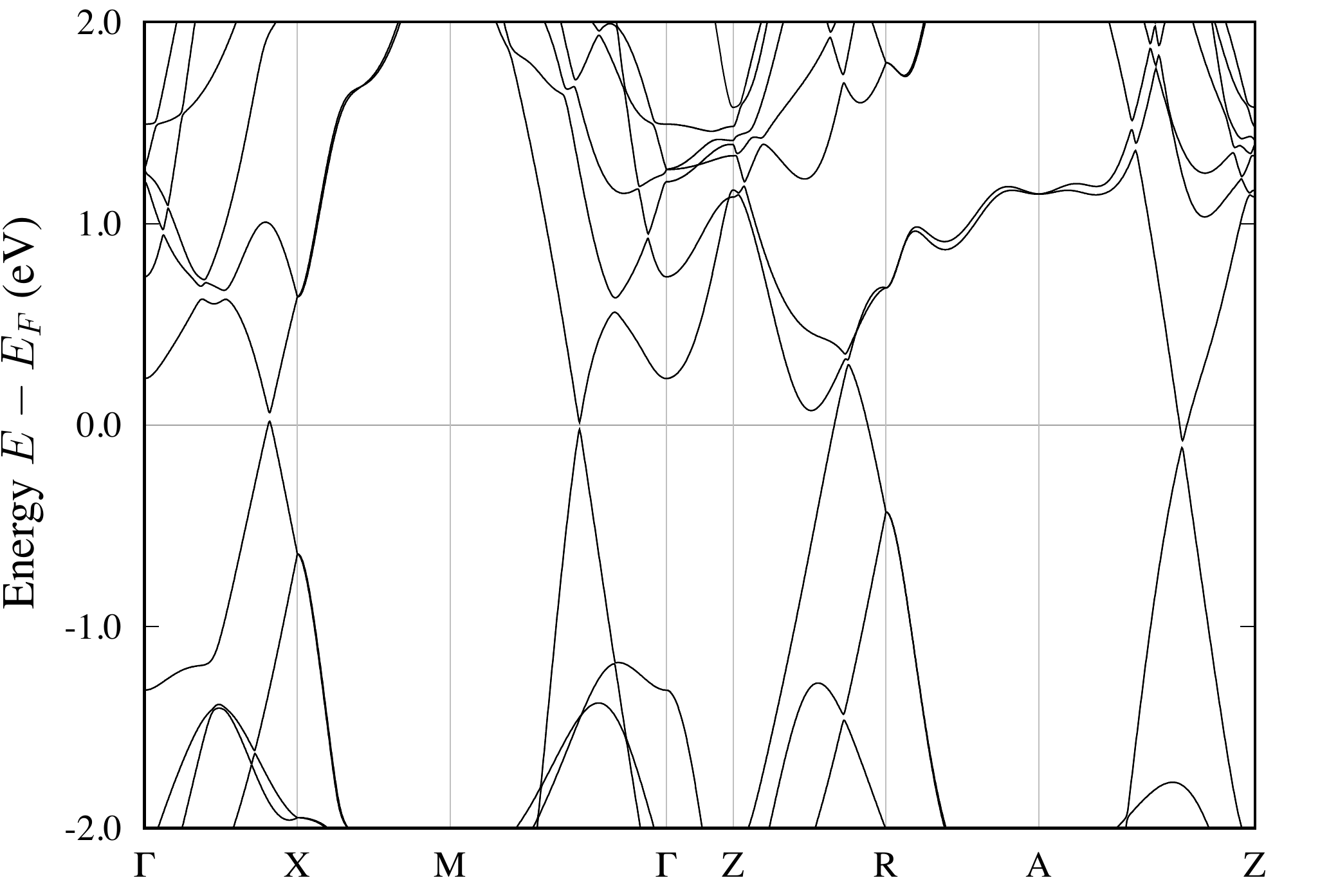}
	\caption{Calculated bandstructure of ZrSiS including spin-orbit coupling, which opens up small gap of $\sim 35\,meV$ for the Dirac cones near the Fermi surface.}
	\label{f2}
\end{figure}

\begin{figure}[h!]
	\centering
		\includegraphics[width=0.7\textwidth]{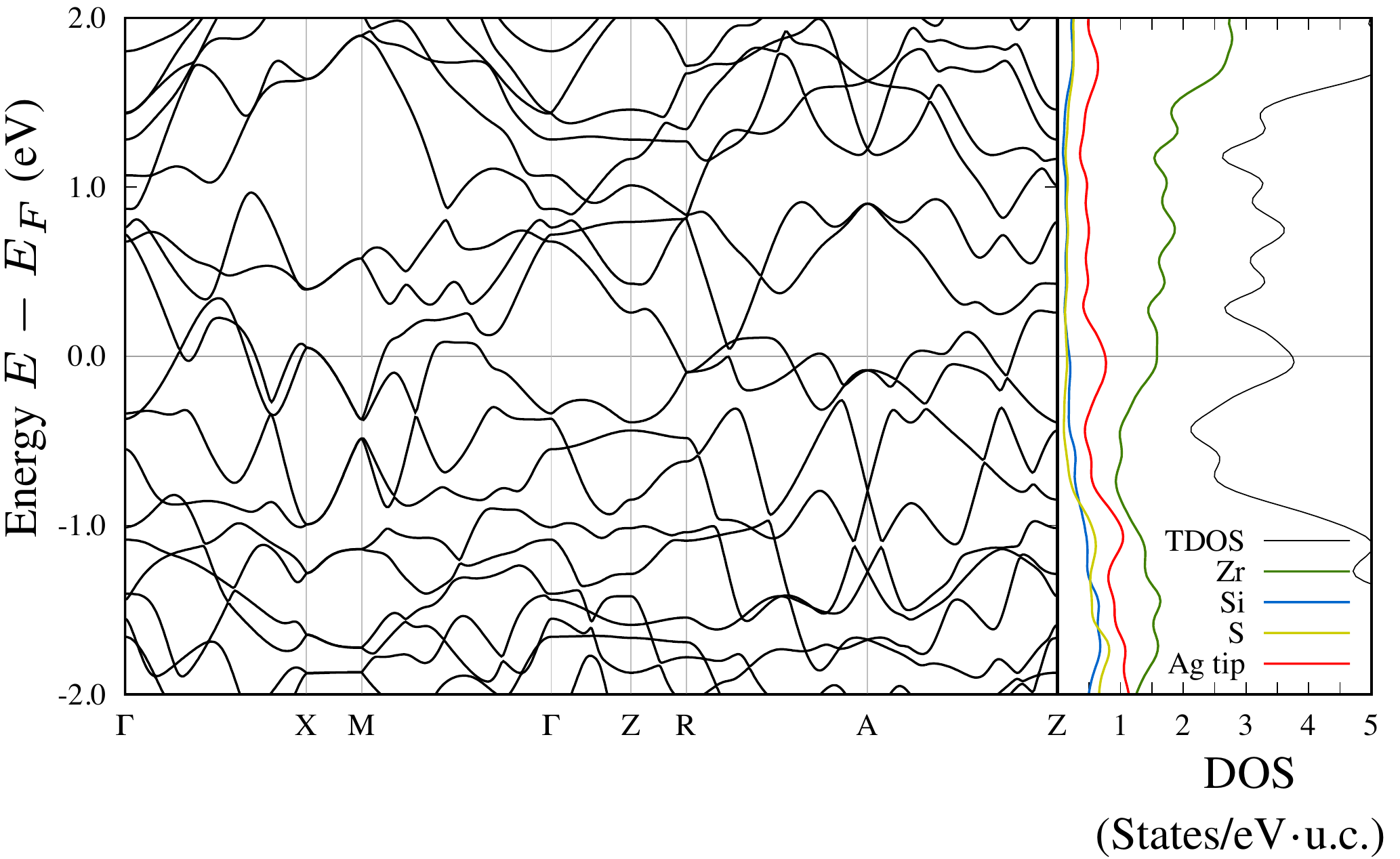}
	\caption{Calculated bandstructure for the 3$\times$1$\times$1 [ZrSiS]$_4$Ag$_6$ superlattice indicating destruction of all the Dirac points.}
	\label{f2}
\end{figure}

\end{document}